\documentclass[prl,aps,superscriptaddress,twocolumn,letter,nopacs,footinbib,longbibliography]{revtex4-2}

\usepackage{array,multirow,graphicx,color,amsmath,amsfonts,enumerate,amsthm,amssymb,mathtools,enumitem,thmtools,hyperref,subfigure,mathdots,enumitem,centernot,bm,soul,bbm,tikz,pgfplots}
\usepackage{enumitem}
\usepackage[capitalise, noabbrev]{cleveref}
\usetikzlibrary{arrows}
\pgfplotsset{compat=1.14}
\hypersetup{colorlinks=true,linkcolor=blue,citecolor=blue,urlcolor=blue}

\usepackage{tcolorbox}\tcbset{before skip=10pt,toptitle=2mm,bottomtitle=1mm,fonttitle=\bfseries}
\tcbuselibrary{theorems}
\tcbuselibrary{breakable}

\definecolor{niceRed}{rgb}{0.825, 0.248, 0.248}
\definecolor{niceBlue}{rgb}{0.248, 0.248, 0.825}
\definecolor{niceGreen}{rgb}{0.125, 0.6, 0.125}

\def\H{ {\cal H} }

\def\L{ {\cal L} }

\def\T{ {\cal T} }

\def\>{\rangle}
\def\<{\langle}
\newcommand{\bra}[1]{\langle {#1} |}
\newcommand{\ket}[1]{| {#1} \rangle}

\newcommand{\tr}[1]{\mathrm{Tr}\left( #1 \right)}

\let\vv\v
\renewcommand{\v}[1]{\ensuremath{\boldsymbol #1}}

\theoremstyle{plain}
\newtheorem{thm}{Theorem}

\theoremstyle{definition}

\newcolumntype{C}[1]{>{\centering\arraybackslash}p{#1}}

\DeclareFontFamily{U}{mathb}{\hyphenchar\font45}
\DeclareFontShape{U}{mathb}{m}{n}{
	<-6> mathb5 <6-7> mathb6 <7-8> mathb7
	<8-9> mathb8 <9-10> mathb9
	<10-12> mathb10 <12-> mathb12
}{}
\DeclareSymbolFont{mathb}{U}{mathb}{m}{n}
\DeclareMathSymbol{\llcurly}{\mathrel}{mathb}{"CE}
\DeclareMathSymbol{\ggcurly}{\mathrel}{mathb}{"CF}

\renewcommand*{\thefootnote}{\fnsymbol{footnote}}

\begin{document}

\title{Optimizing thermalizations}

\author{Kamil Korzekwa}
\affiliation{Faculty of Physics, Astronomy and Applied Computer Science, Jagiellonian University, 30-348 Krak\'{o}w, Poland}

\author{Matteo Lostaglio}
\affiliation{Korteweg-de Vries Institute for Mathematics and QuSoft, University of Amsterdam, The Netherlands}
\affiliation{QuTech, Delft University of Technology, P.O. Box 5046, 2600 GA Delft, The Netherlands}

%\footnotetext{These authors contributed equally to this work. \\ Emails: lostaglio@protonmail.com, korzekwa.kamil@gmail.com}

\begin{abstract}
	We present a rigorous approach, based on the concept of continuous thermomajorisation, to algorithmically characterise the full set of energy occupations of a quantum system accessible from a given initial state through weak interactions with a heat bath. The algorithm can be deployed to solve complex optimization problems in out-of-equilibrium setups and it returns explicit elementary control sequences realizing optimal transformations. We illustrate this by finding optimal protocols in the context of cooling, work extraction and catalysis. The same tools also allow one to quantitatively assess the role played by memory effects in the performance of thermodynamic protocols. We obtained exhaustive solutions on a laptop machine for systems with dimension $d\leq 7$, but with heuristic methods one could access much higher $d$.
\end{abstract}

\maketitle
\renewcommand{\thefootnote}{\arabic{footnote}}
\setcounter{footnote}{0}

%------------------------------------------------------------------
% INTRODUCTION
%------------------------------------------------------------------

\paragraph{Introduction.} Thermalization is a ubiquitous process across quantum sciences. A control over interactions between a quantum system and its thermal environment is pivotal for applications ranging from quantum computing to thermodynamics and, quite broadly, for demonstrating and harnessing quantum effects in a variety of experimental platforms. Algorithmic cooling protocols initializing pure ancilla states~\cite{schulman2005physical,rodriguez2020novel}, quantum heat engines~\cite{curzon1975efficiency,esposito2009universality,uzdin2014universal}, fridges and heat pumps~\cite{segal2006molecular,allahverdyan2010optimal,correa2013performance}, dissipative generation of quantum entanglement~\cite{brask2015autonomous} -- these are but a few examples of tasks involving thermalization stages. These all require the identification of optimal thermalization schedules maximizing corresponding objective functions (ground state population, efficiency, cooling power, entanglement, etc.). 

Performing such optimizations in full generality is typically hopeless. In fact, thermalizations are routinely described by quantum Markovian master equations~\cite{lindblad1976generators}, and the problem of characterising all dynamics that can result from the integration of such a class of differential equations is extremely challenging even classically, where it is known as the \emph{embeddability problem}~\cite{elfving1937theorie,davies2010embeddable}. Despite having been studied for decades in the mathematics literature, general solutions are not known beyond the simplest cases of dimension $d=2$ and $d=3$~\cite{kingman1962imbedding,runnenburg1962elfving,goodman1970intrinsic,carette1995characterizations}. 

Here, we start from the observation that the following properties hold in wide generality across standard thermalization models used in atomic, molecular and optical physics, quantum thermodynamics and quantum computing~\cite{breuer2002open,kosloff2013quantum, alicki2018introduction,dann2021open}:
\begin{enumerate}[label=(P\arabic*),itemsep=0mm]
	\item \label{p1} The evolution of the system's quantum state is described by a Markovian master equation.
	\item \label{p2} The thermal state is a fixed point of the evolution.
	\item \label{p3} Evolutions of the energy occupations (``populations'') and that of energetic quantum superpositions (``coherences'') decouple. 
\end{enumerate}  
There are numerous examples of phenomena described by equations satisfying the above assumptions: thermalization strokes in thermal machines~\cite{uzdin2015equivalence, levy2018quantum}, weak thermal contact of a quantum system with a large environment~\cite{roga2010davies, lindblad1976generators}, an atomic degree of freedom interacting weakly with a thermal radiation field~\cite{breuer2002open}, depolarization noise in a quantum computer and strong coupling scenarios via reaction coordinates~\cite{nazir2018reaction}. For brevity, we will refer to any such dynamical thermalization model as a \emph{Markovian Thermal Process} or an MTP.

The questions we address in this work are: 
\begin{enumerate}[label=(Q\arabic*),itemsep=0mm]
	\item\label{q1} What are the most general constraints characterising the evolution of energy occupations (populations) under an MTP?
	\item\label{q2} Given an initial out-of-equilibrium population, can we construct the set of all populations achievable from it under arbitrary MTP?
	\item\label{q3} 	If a transformation from an initial to a final population is possible under MTPs, can we construct an explicit control sequence realizing it?
	
\end{enumerate}  
In what follows we employ the mathematical framework developed in the accompanying paper~\cite{lostaglio2022continuous} to provide answers to all these questions. It turns out that~\ref{q1} corresponds to a broad generalization of the well-known entropy production inequalities~\cite{landi2021irreversible}. Concerning~\ref{q2}, we answer the question in the affirmative by providing an explicit verification algorithm and offering its Mathematica implementation~\cite{korzekwa2021continuous}. Here, we shall illustrate how this algorithm can be used to solve optimization problems involving cooling, work extraction and catalysis. We also report on a recent work which used our results to show that non-Markovianity boosts the efficiency of thermal bio-molecular switches~\cite{spaventa2021non}. Finally, in answering~\ref{q3}, we pave the way to the idea of algorithmically generated thermodynamic protocols, as we showcase by using our code to generate the optimal cooling protocol recently introduced in Ref.~\cite{rodriguez2017heat}. These applications build up encouraging evidence that the framework is suitable to perform model-independent optimizations, as well as to algorithmically construct new thermodynamic schemes in regimes where a complete analysis is unattainable by alternative methods. 

%------------------------------------------------------------------
% MOTIVATING EXAMPLE
%------------------------------------------------------------------

\paragraph{Motivating example.} Before we go into the details of the general framework, we illustrate the idea within an experimentally relevant setting. Consider Heat-Bath Algorithmic Cooling (HBAC) protocols~\cite{schulman1999molecular,boykin2002algorithmic,rodriguez2016achievable, raeisi2015asymptotic, rodriguez2020novel}, whose aim is to achieve the largest possible cooling of a target system by means of protocols involving two main steps. First, a unitary interaction can be performed that involves the target system and some thermal ancilla qubits. Second, there is an interaction between system+ancillas and a thermal environment. The two steps are repeated several times. 

Experimental realizations have been demonstrated in the context of (liquid and solid) state NMR, ion traps and quantum optical setups, among others~\cite{rodriguez2020novel}. In the well-known PPA protocol~\cite{schulman2005physical}, the interaction with the environment simply resets the ancilla qubits to a thermal state. For a long time, this was the best known scheme. More recently, the state-reset-$\Gamma$ (SR$\Gamma$) protocol~\cite{rodriguez2017heat} was introduced. This is based on the Nuclear Overhauser Effect, and performs a reset of the submanifold $\{\ket{00 \dots 0}$, $\ket{11 \dots 1}\}$ of the energy levels of system+ancillas, hence involving collective interactions which lead to better cooling. Note that in both protocols the interaction with the environment is described by an MTP. 

We ask: is the SR$\Gamma$ optimal, or could even better cooling be achieved with similar level of control by tweaking the thermalization dynamics? Our framework algorithmically returns the SR$\Gamma$ protocol as the best available algorithmic cooling protocol on a qubit target system and a qubit thermal ancilla, where the optimization is carried out over all MTPs~\footnote{For details, see Supplemental Material that includes Refs.~\cite{marvianthesis,lostaglio2015description,halpern2020fundamental,preliminary}}. This is not only a previously unknown optimality result, but it also showcases how one can move from guesswork to optimization in devising cooling protocols.  

%------------------------------------------------------------------
% SETTING
%------------------------------------------------------------------

\emph{Setting.} Formalizing the discussion in the introduction, an MTP is any dynamics generated by a Markovian master equation (due to~\ref{p1}),
\begin{equation}
	\label{eq:master}
	\frac{d \rho(t)}{dt}  = \H(\rho(t)) + \mathcal{L}_t(\rho(t)),
\end{equation}
where $\H(\cdot)=- i [H, \rho]$ is the generator of a closed, reversible quantum dynamics, with $H$ being the (potentially dressed) Hamiltonian of the system, and $\mathcal{L}_t$ being a Lindbladian generating an open, irreversible quantum dynamics~\cite{gorini1976completely,lindblad1976generators}. Moreover, due to~\ref{p2}, the Gibbs thermal state \mbox{$\gamma = e^{-\beta H}/\tr{e^{-\beta H}}$} is the stationary solution, $\L_t\gamma=0$. Due to~\ref{p3}, $\L_t$ commutes with $\H$ at all times $t$. Crucially, typical microscopic derivations employing the weak coupling limit~\cite{breuer2002open,kosloff2013quantum, alicki2018introduction} lead to a master equation with this general form. For example, Davies maps~\cite{davies1974markovian, roga2010davies} describing the thermalization with a large bath in the weak coupling limit are of this form, with $\mathcal{L}_t$ independent of $t$. Another example is given by each stroke of a two-stroke engine cycle in the weak coupling limit, since the time dependence of $H$ disappears in interaction picture~\cite{uzdin2015equivalence}. Hence, the class of equations specified by Eq.~\eqref{eq:master} emerges in many situations after the relevant approximations.

What is more, in the accompanying paper~\cite{lostaglio2022continuous}, we show that \emph{every} transformation achievable under \emph{arbitrary} MTP can be also achieved within a straightforward physical setup:
\begin{thm}
	\label{thm:elementary}
	Any transformation achievable by an MTP can be realized by a sequence of elementary thermalizations, each involving only $2$ energy levels of the system.
\end{thm}
Elementary thermalizations between two levels $(i,j)$ are simply described by a standard reset master equation:
\begin{subequations}
	\begin{align}
		\frac{dp_i(t)}{dt}&=\frac{1}{\tau}\left(  \frac{\gamma_i}{\gamma_i+\gamma_j}(p_i(t)+p_j(t))-p_i(t) \right),
	\end{align}
\end{subequations}
and an analogous one exchanging $i$ and $j$. Here and in what follows, $\v{p}(t)$ is a vector of populations with $p_i(t) = \bra{E_i} \rho(t) \ket{E_i}$, where $\ket{E_i}$ is the eigenstate of $H$ corresponding to energy $E_i$ (for simplicity we consider a non-degenerate $H$), while $\v{\gamma}$ is a vector of thermal populations. Elementary thermalizations describe an exponential relaxations of energy levels $(i,j)$ to a thermal state. For example, one may have to first thermalize levels $1$ and $3$ for some time, then $2$ and $5$, etc.  

Such elementary thermalizations arise naturally from the weak interaction of a two-level system with a large bath~\cite{davies1974markovian, roga2010davies} and collision models~\cite{scarani2002thermalizing}. In practice, they are often used as building blocks for more complex protocols~\cite{perry2018sufficient}, in the context of work extraction~\cite{baumer2019imperfect} and slow driving~\cite{miller2019work}. Theorem~\ref{thm:elementary} states that elementary thermalizations are a universal set of controls for MTP (in contrast to the non-Markovian regime, where it is not the case~\cite{lostaglio2018elementary}). 

The ability to selectively couple certain energy levels to the bath (ideally switching instantaneously from one to the other) is a commonplace assumption in the study of discrete engines, when one can couple and decouple submanifolds of the system's energy spectrum~\footnote{Typically one works in interaction picture and ignores a (small) Lamb shift.} with baths at different temperatures~\cite{uzdin2015equivalence, levy2018quantum}. Note, however, that two-level thermalizations on multi-qubit systems require highly non-local interactions and hence may be challenging to implement. Also, we highlight that generally there will be infinitely many other protocols connecting the same two states. As we shall see, it is convenient to focus on elementary thermalizations because the necessary controls can be algorithmically constructed. In any case, we study thermalization independently of whether and how this stage is embedded in a more complex protocol. 

%------------------------------------------------------------------
% GENERALIZED ENTROPY PRODUCTION INEQUALITIES
%------------------------------------------------------------------

\paragraph{Generalized entropy production inequalities (GEP).} In the accompanying paper~\cite{lostaglio2022continuous}, we derive the most general set of constrains that need to be satisfied by populations $\v{p}(t)$ independently of the details of the MTP generating the evolution. These conditions subsume (and greatly strengthen) the standard positive entropy production inequality
\begin{equation}
	\label{eq:standard_entropy_prod}
	\frac{d \Sigma(t)}{dt} =	\frac{dS(t)}{dt} - \beta J(t) \geq 0,  
\end{equation}
where \mbox{$S(t) := - \tr{\rho(t) \log \rho(t)}$} is the von Neumann entropy and \mbox{$J := \tr{H d\rho(t)/dt}$} is the heat current flowing to the system. Specifically,  for any well-behaved convex function \mbox{$h: \mathbb{R} \rightarrow \mathbb{R}$}, the $h$-divergence defined by
\begin{equation}
	\label{eq:h_divergence}
	\Sigma_h(t) = - \sum_{i=1}^d \gamma_i h\left( \frac{p_i(t)}{\gamma_i} \right),
\end{equation}
must be monotonically non-decreasing, $d\Sigma_h(t)/dt \geq 0$. For each choice of $h$, the above qualifies as a valid GEP inequality.

Choosing $h(x)=x \log(x)$, one obtains: $d \Sigma_{\rm d}(t)/dt = d S_{\rm d}(t)/dt - \beta J(t) \geq 0$, where $S_{\rm d}(t) = - \sum_i p_i(t) \log p_i(t)$ is often referred to as the \emph{diagonal entropy} \cite{polkovnikov2011microscopic}. This recovers a result recently appearing in Ref.~\cite{santos2019role}, and can be used to derive the standard entropy production inequality in Eq.~\eqref{eq:standard_entropy_prod}~\cite{Note1}. Hence, the GEP framework implies the usual entropy production relation. Choosing $h(x) = {\rm sgn}(\alpha) x^\alpha/(\alpha-1)$ for $\alpha\in\mathbb{R}$, one obtains $d\Sigma_\alpha(t)/dt\geq 0$, where $\Sigma_\alpha(t) = - S_\alpha(\v{p}(t)\| \v{\gamma})$ is the relative R\'enyi entropy. This recovers the ``second laws'' of Ref.~\cite{brandao2013second} in a more stringent form. In contrast to the end-point conditions of Ref.~\cite{brandao2013second}, our constraints prescribe that all the $\Sigma_\alpha$ must be constantly non-decreasing along the dynamical trajectory $\v{p}(t)$. When $\v{\gamma}$ is a uniform distribution (infinite temperature limit) the above conditions reduce to the non-decreasing of all R\'enyi entropies~\cite{renyi1961measures}. Another relevant class of GEP inequalities can be found taking $h_q(x) = \textrm{sgn}(q) (1- x^q)/(1-q) $, giving $d\Sigma^T_q(t)/dt \geq 0$, with $\Sigma^T_q(t) = - S^T_q(\v{p}(t)\| \v{\gamma})$ and $S^T_q(\v{p}\| \v{\gamma}) :=  \textrm{sgn}(q)( \sum_i p_i^q \gamma_i^{1-q} -1)/(q-1)$ being the Tsallis relative entropy. The Tsallis entropies, well-known in non-extensive statistical mechanics and information theory~\cite{tsallis1988possible, abe2000axioms, tsallis2009introduction}, are recovered in the infinite temperature limit. This extends to arbitrary temperatures the results of Ref.~\cite{mariz1992irreversible}. Taking $h(x)= - \log x$ we get $- d \mathcal{V}(t)/dt \geq 0$, with $\mathcal{V}(t) = - S(\v{\gamma}\| \v{p}(t))$ the `vacancy', found in Ref.~\cite{wilming2017third} to be the central constraint at very low temperatures. 

GEP inequalities encompass a wealth of disparate results as part of a unified framework. At the same time, a natural question arises: Is there a family of entropic conditions that implies \emph{all others}? The answer is affirmative and can be interpreted as a sort of exhaustive $H$-theorem (for the proof see the accompanying paper~\cite{lostaglio2022continuous}).
\begin{thm}
	\label{thm:gen_entropy_prod}
	All GEP are implied by the non-decreasing of
	\begin{equation}
		\label{eq:generalisedentropicrelations}
		\Sigma_a(t) := -\sum_{i=1}^d \left|p_i(t) - a \frac{\gamma_i}{\gamma_d}\right|, \quad a \in [0,1].
	\end{equation}
 What is more, if there exists a trajectory connecting $\v{p}(0)$ and $\v{p}(t_f)$ along which all $\Sigma_a$ do not decrease, then there exists an MTP transforming $\v{p}(0)$ into $\v{p}(t_f)$.
\end{thm}
Not only $\Sigma_a$ are a complete set of GEP, but they even guarantee the existence of a physical realization. In contrast, to satisfy the standard entropy production relation along a trajectory does not ensure the existence of a physical process.

%------------------------------------------------------------------
% ALGORITHMIC VERIFICATION AND CONSTRUCTION
%------------------------------------------------------------------

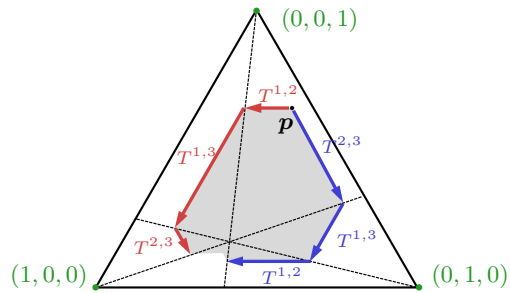
\begin{figure}[t]
	\centering
	\begin{tikzpicture}
	% External picture
	\node (myfirstpic) at (0,0) {\includegraphics[width=\columnwidth]{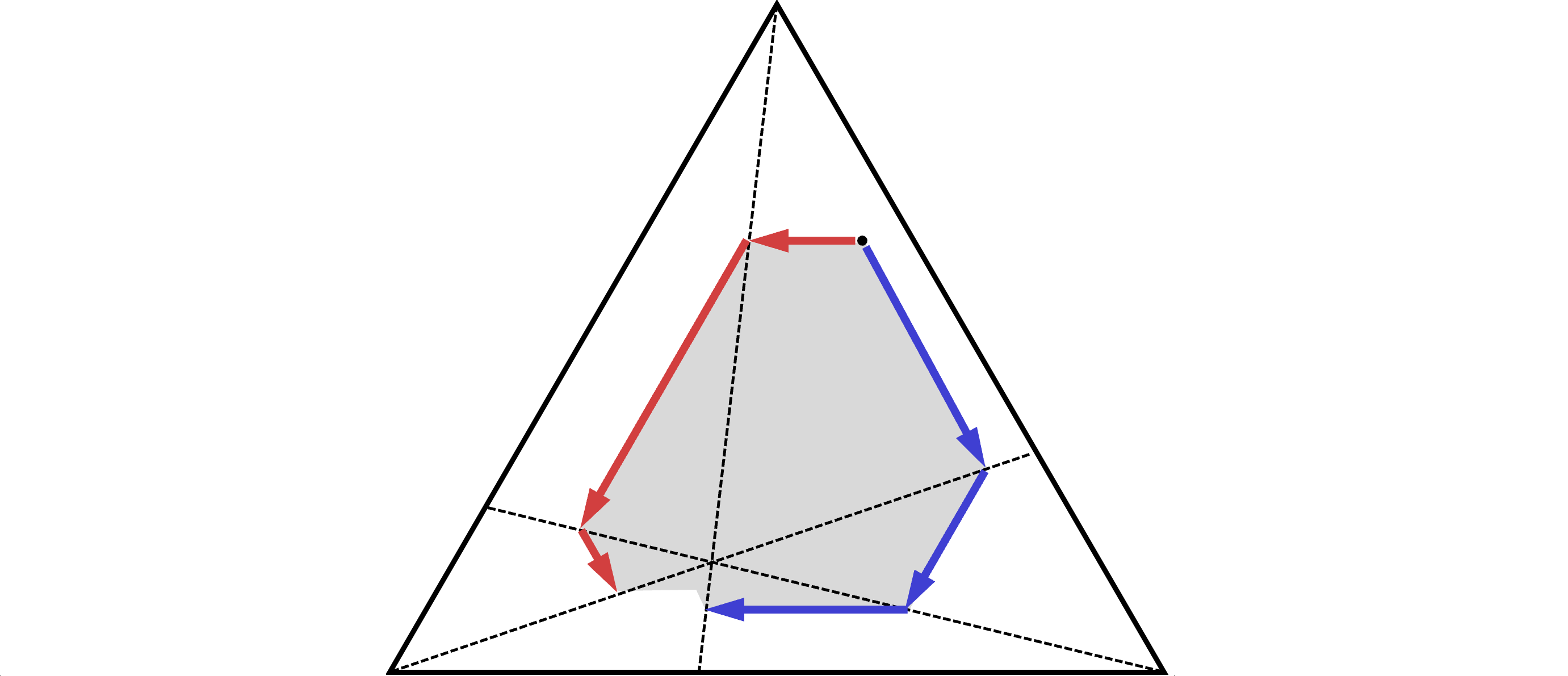}};
	% Labels
	%\node at (-3.5,2.5) {(b)};
	\node at (0.35,0.3) {$\v{p}$};
	\node at (1.1,0.05) {\scriptsize{{\color{niceBlue}$T^{2,3}$}}};
	\node at (0.3,-1.7) {\scriptsize{{\color{niceBlue}$T^{1,2}$}}};
	\node at (1.3,-1.2) {\scriptsize{{\color{niceBlue}$T^{1,3}$}}};
	\node at (0.25,0.73) {\scriptsize{{\color{niceRed}$T^{1,2}$}}};
	\node at (-1.4 ,-1.3) {\scriptsize{{\color{niceRed}$T^{2,3}$}}};
	\node at (-0.83,-0.1) {\scriptsize{{\color{niceRed}$T^{1,3}$}}};
	\node at (2.25,-2.3) {\phantom{$(0,1,0)$}};
	
	\node  at (-2.8,-1.7) {{\color{niceGreen}$(1,0,0)$}};
	\node at (2.8,-1.7) {{\color{niceGreen}$(0,1,0)$}};
	\node at (0.8,1.7) {{\color{niceGreen}$(0,0,1)$}};
	
	\filldraw[niceGreen] (-0.04,1.83) circle (1pt);
	\filldraw[niceGreen] (2.12,-1.85) circle (1pt);
	\filldraw[niceGreen] (-2.18,-1.85) circle (1pt);
	\end{tikzpicture}
	\caption{\label{fig:algorithm}\textbf{Markovian thermalization in $d=3$.} Simplex representing the state space of all 3-dimensional probability distributions. The grey area denotes the set of achievable states from $\v{p}$ under arbitrary Markovian Thermal Processes when the thermal state is $\v{\gamma}=[1/2,1/3,1/6]$. The elementary thermalizations $T^{i,j}$, which fully thermalize the pair of levels $i$ and $j$  and whose action is indicated by red and blue arrows, play a central role in the sequence of required controls. }
\end{figure}

\emph{Algorithmic verification and construction.} If we could construct the set of states $\T_+(\v{p}(0))$ achievable under MTPs from a given $\v{p}(0)$, we could use this knowledge for our optimization purposes. GEP inequalities encode such information, but of course one cannot check an infinite number of conditions along an infinite number of potential trajectories. Furthermore, they carry no information about the required controls. Remarkably, both problems can be solved in full generality employing the notion of continuous thermomajorization, as we explain in detail in the accompanying paper~\cite{lostaglio2022continuous}.

\begin{thm}
	Given $\v{p}(0)$ the set of all states $\T_+(\v{p}(0))$ achievable from it via arbitrary MTPs can be constructed by an algorithm in a finite number of steps. Moreover, the algorithm outputs a sequence of elementary thermalizations realizing any transformation of interest.
\end{thm} 

We provide a corresponding Mathematica code in~\cite{korzekwa2021continuous}, and present a toy example for a system of dimension $d=3$ in Fig.~\ref{fig:algorithm}. On a laptop machine, the algorithm solves the $d=6$ case in minutes and $d=7$ case in hours. Thus, our framework allows one to transform an intractable optimization problem for an objective function $g$ into an explicit form:
\begin{equation}
	\max_{\{\v{p}(t_f)\}}g(\v{p}(t_f)) = \max_{\v{q}\in \T_+(\v{p}(0))}  g(\v{q}),
\end{equation}
where $\v{p}(t_f)$ are all probability distributions achievable from $\v{p}(0)$ via MTPs. Whenever $g$ is convex, the new optimization problem is finitely verifiable, because the algorithm returns all extremal points of the set $\T_+(\v{p}(0))$. Note, however, that the set $\T_+(\v{p}(0))$ is not generally convex (as can be seen, e.g., in Fig.~\ref{fig:algorithm}).

%------------------------------------------------------------------
% APPLICATIONS
%------------------------------------------------------------------

\paragraph{Applications.} We now proceed to illustrating some applications of our framework (see~\cite{Note1} for further details). The aim is to provide evidence of its usefulness, while we believe much more can be done at the interface with open quantum system dynamics. 

We start off with the fundamental protocol of work extraction. Work in quantum thermodynamics is often treated as a random variable, but for microscopic systems its average can be of the same order of magnitude as its fluctuations~\cite{aberg2013truly}. Hence, much attention has been given to taming such fluctuations~\cite{horodecki2013fundamental,aberg2013truly, richens2016work, perarnau2019collective}. Single-shot work extraction protocols guarantee to extract an amount of work $W$ up to a failure probability $\epsilon$. More precisely, given an out of equilibrium system $S$ and access to a thermal environment $E$ at inverse temperature $\beta=1/(kT)$, the task is to deterministically excite a battery system $B$, initially prepared in an energy eigenstate $E_0$, over an energy gap $W$. When the probability that $B$ has energy $E_0 +W$ is at least $1-\epsilon$, one extracts $\epsilon$-deterministic work $W$~\cite{horodecki2013fundamental}. 

\begin{figure}
	\includegraphics[width=0.65\columnwidth]{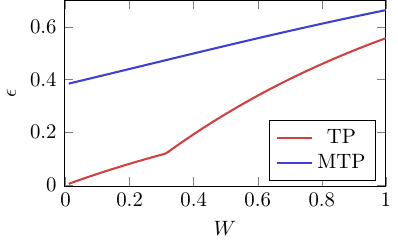}
	\caption{\label{fig:work}\textbf{$\epsilon$-deterministic work extraction from a two-level system.} Minimal error $\epsilon$ as a function of the work $W$ extracted from a two-level system with energy splitting $\Delta$ prepared in a thermal state at temperature $1/\beta_S$ smaller than the environmental temperature $1/\beta_E$. System-environment interactions are modelled by TP or MTP. Parameters: \mbox{$\beta_S \Delta=2$} and \mbox{$\beta_E \Delta=1$}. }
\end{figure}

The optimal extraction error under Thermal Processes (TP), i.e. all processes satisfying~\ref{p2}-\ref{p3} but not necessarily~\ref{p1} (so possibly non-Markovian), can be computed via the thermomajorization condition of Ref.~\cite{horodecki2013fundamental}. However, when one is limited to MTP, the optimal error $\epsilon_{\mathrm{MTP}}(W)$ can be computed with our algorithm. We observe that memory effects dramatically decrease the minimal error for given $W$, see Fig.~\ref{fig:work}. Note that $\epsilon_{\mathrm{MTP}}(W)$ remains very high even in the $W \rightarrow 0$ limit, showing that converting a non-equilibrium resource into deterministic work either requires control over the energy spectrum (as in Ref.~\cite{aberg2013truly, perry2018sufficient}), or otherwise relies on environmental memory effects.

\begin{figure}[t]
	\includegraphics[width=0.8\columnwidth]{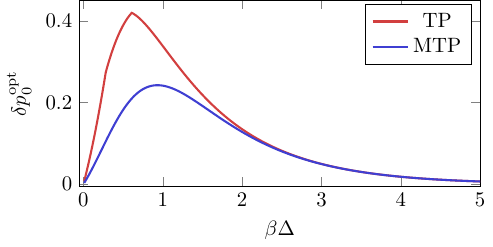}
	\caption{\label{fig:cooling}\textbf{Optimal cooling of a four-level system.} Optimal change of the ground state population $\delta p_0^{\mathrm{opt}}$ for a four-level quantum system initially at equilibrium with inverse temperature $\beta$ in one round of algorithmic cooling consisting of a unitary inversion of the populations, followed by the optimal TP or MTP. The system has equidistant energy spectrum with smallest energy difference $\Delta$.  }
\end{figure}

In a similar manner, we can provide model-independent evidence of the role played by system-environment correlations in boosting cooling processes. Consider a system initially at equilibrium with the environment at inverse temperature $\beta$, and the task of cooling it down by maximising its ground state occupation through a round of algorithmic cooling protocol. The first step of the protocol is to unitarily invert the occupations (while thermal occupations are monotonically decreasing with growing energy, the inverted occupations are monotonically increasing); and the second step is to optimally interact it with the bath to maximise the occupation of the ground state. Now, the interactions may be optimised over all TPs or all MTPs, and the difference in performance quantifies the boost to cooling due to memory effects. We illustrate this in Fig.~\ref{fig:cooling} for a four-level system with an equidistant spectrum.

Our framework allows one not only to investigate the difference in performance of thermodynamic protocols with and without memory, but also to interpolate between these two extremes. This question relates to the well-known topic of \emph{catalysis} in thermodynamics~\cite{brandao2013second, ng2015limits, lipka2021all}. Catalysis is the phenomenon by which a certain transformation $\v{p}(0) \mapsto \v{q}$ is only possible when aided by an auxiliary system $\v{c}$ (the catalyst, playing the role of a memory), which is given back unchanged and uncorrelated at the end: $\v{p}(0) \otimes \v{c} \mapsto \v{q} \otimes \v{c}$. To illustrate this, we consider a two-level system initially at temperature two times higher than that of the bath. Under MTPs and with no catalyst, the system can be cooled at most to the temperature of the bath~\cite{alhambra2019heat,wilming2016second}. Our algorithm shows that a qubit thermal catalyst already allows one to cool the original system below the bath's temperature, highlighting how catalysis is useful not only in the abstract resource theory setting \cite{brandao2013second, ng2015limits, lipka2021all}, but also in the context of Markovian master equations describing standard thermalization models. What is more, differently from previous approaches, the algorithm returns an explicit set of controls implementing the protocol. This tackles one of the main challenges that insofar prevented the general results on catalysis to impact practical protocols.

Finally, we want to point out that our our framework was recently used to obtain rigorous, quantitative evidence that environmental memory effects are beneficial to the efficiency of bio-molecular processes~\cite{spaventa2021non}. More precisely, the authors of~\cite{spaventa2021non} demonstrated that the optimal photoisomerization yield in the presence of memory effects is strictly larger than in the Markovian regime.

%------------------------------------------------------------------
% OUTLOOK
%------------------------------------------------------------------

\paragraph{Outlook.} We constructed a framework to systematically perform optimization over thermalization processes and algorithmically construct optimal protocols. The algorithm yielded encouraging results in a range of relevant applications. A promising direction to probe higher dimensional systems ($d \geq 8$) is to relax the current exhaustive search to a heuristic algorithm, or to focus on short elementary thermalization sequences. Pushing the achievable dimension up, and combining the current algorithm optimizing the thermalization stage with alternative methods to optimize unitary stages (as we implicitly did in the HBAC application), will likely open up a range of applications, such as the optimisation of quantum thermodynamic cycles of heat engines~\footnote{Here is a naive proposal. In a four-stroke engine, one can apply our algorithm to each of the thermalization strokes, computing all available final states, while unitary strokes are tackled by the usual methods. Using the elementary thermalization model, one then associates to each final state a time required to achieve it. This allows one to optimize functions of power and efficiency.}. At the same time, our framework also offers a rigorous information-theoretical foundation to the dynamical viewpoint of quantum thermodynamics complementing the toolbox of master equation approaches, as we have seen with the systematic construction of GEP inequalities.

\bigskip

\begin{acknowledgments}
\noindent \textbf{Acknowledgments.} Both authors contributed equally to this work. M.L. thanks A. Levy for useful discussions. K.K. acknowledges financial support by the Foundation for Polish Science through TEAM-NET project (contract no. POIR.04.04.00-00-17C1/18-00). M.L. acknowledges financial support from the the European Union's Marie Sk{\l}odowska-Curie individual Fellowships (H2020-MSCA-IF-2017, GA794842), Spanish MINECO (Severo Ochoa SEV-2015-0522 and project QIBEQI FIS2016-80773-P), Fundacio Cellex and Generalitat de Catalunya (CERCA Programme and SGR 875) and grant EQEC No. 682726.
\end{acknowledgments}

\bibliography{Bibliography}

\clearpage
\setcounter{equation}{0}
\setcounter{figure}{0}
\renewcommand{\theequation}{S\arabic{equation}}
\renewcommand{\thefigure}{S\arabic{figure}}

\onecolumngrid
\widetext
\begin{center}
	\textbf{\large Supplemental Material}
\end{center}

\section{The SR$\Gamma$ cooling protocol: algorithmic discovery and extrapolation}

Since only recently the SR$\Gamma$ protocol was discovered~\cite{rodriguez2017heat}, and the thermalization step is simply postulated, it is left open whether better cooling protocols with same controls exist. Also, it is natural to ask whether our algorithm can generate the SR$\Gamma$ protocol as the result of an optimization process rather than guesswork. Here, we illustrate the power of our framework by showing how our algorithm returns the SR$\Gamma$ protocol as the best available cooling protocol on a qubit target system, when using a single qubit thermal ancilla and arbitrary Markovian interactions with the thermal environment.   

Let us formalize the question. Let $\v{p}^{(k)}$ be the output state of the target system after $k$ rounds of a generic HBAC protocol. Step $k+1$ of the cooling protocol consists of (a) performing a unitary process on target+ancilla \mbox{$\v{p}^{(k)} \otimes \v{\gamma} \stackrel{U}{\mapsto} \v{s}^{(k)}$}; (b) finding an optimal Markovian thermal process
\begin{equation}
	\v{s}^{(k)} \stackrel{ \textrm{MTP}}{\longmapsto} 	\v{p}^{(k+1)} \otimes \v{\gamma},
\end{equation}
where $\v{\gamma}$ is the thermal ancilla qubit that is given back unchanged and $\v{p}^{(k+1)}$ is the output state which maximises the ground state population over all MTPs.

That the optimal unitary must be a permutation follows from the fact that the objective function is linear. For each permutation, we use our algorithm to find the optimal MTP. We find that:
\begin{itemize}
	\item There are $5$ different choices of $U$ that lead to optimal cooling. One of them is to simply to apply a Pauli $X$ on the thermal ancilla.
	\item For each of the $5$ choices of $U$, there is an optimal Markovian thermalization. In the case where \mbox{$U=X$}, the optimal thermalization involves first thermalizing the subspace \mbox{$\{\ket{00}$, $\ket{11}\}$} and then re-thermalizing the ancilla.
\end{itemize}
Remarkably, this means that our protocol returns the SR$\Gamma$ protocol (and $4$ other similar ones) as the optimal HBAC. This illustrates that our algorithm is capable of explicitly constructing useful protocols and proving their optimality. 

Even though the same reasoning cannot be straightforwardly applied to higher dimensional systems, this two qubit result clearly suggests the general structure of SR$\Gamma$ protocol. Lower dimensional optimizations followed by extrapolations may be a convenient strategy to devise new thermodynamic protocols.

\section{Recovery of the standard second law}

Here we show that the second law in the form of the standard entropy production relation, 
\begin{equation}
	\frac{d \Sigma(t)}{dt} =	\frac{dS(t)}{dt} - \beta J(t) \geq 0,  
\end{equation}
where \mbox{$S(t) := - \tr{\rho(t) \log \rho(t)}$} is the von Neumann entropy and \mbox{$J := \tr{H d\rho(t)/dt}$} is the heat current, follows from continuous thermomajorisation and simple considerations about the evolution of quantum coherence during the thermal process. 

The GEP inequalities discussed in the main text include the diagonal entropy production:
\begin{equation}
	d \Sigma_{\rm d}(t)/dt = d S_{\rm d}(t)/dt - \beta J(t) \geq 0,
\end{equation}
where $S_{\rm d}(t) = - \sum_i p_i(t) \log p_i(t)$. Now, note that
\begin{equation}
	\Sigma(t) = \Sigma_{\rm d}(t) - A(t), \quad S(t) = S_{\mathrm{d}}(t) - A(t), 
\end{equation}
where $A(t)$ is the \emph{asymmetry} $A(t) := S(\rho(t)\| \rho_{\mathrm{d}}(t))$~\cite{marvianthesis} and $\rho_{\mathrm{d}}$ denotes the fully decohered version of $\rho$ (i.e., block-diagonal state resulting from $\rho$ after setting to zero all terms in the density matrix corresponding to superpositions between different energy states). It is not difficult to show that
\begin{equation}
	-dA(t)/dt \geq 0,
\end{equation}
an inequality measuring the entropic contribution due to loss of quantum coherence~\cite{lostaglio2015description, santos2019role}. This inequality can be simply obtained from the data processing inequality of the quantum relative entropy, property (P3) of MTP (which implies that the dynamics commutes with the decoherence channel) and property (P1) of MTP (which implies the divisibility of the process). 

Starting from the diagonal entropy production inequality and summing the positive contribution  $-d A(t)/dt$, we recover the standard entropy production inequality. Clearly, the two separate relations ($d \Sigma_{\rm d}(t)/dt \geq 0$, $-d A(t)/dt \geq0$) are stronger than the standard entropy production inequality.

\section{Details concerning applications}

\emph{$\epsilon$-deterministic work extraction.} When one has complete control over the interactions between the extended system $SB$ and the bath $E$ (including the environmental memory effects due to the correlations created over time between $SB$ and $E$), the optimal work yield is specified by the thermomajorisation relation. More precisely, the optimal extraction error $\epsilon_{\mathrm{TP}}(W)$ under thermal processes is given by minimal $\epsilon$ satisfying
\begin{equation}
	\v{p}_S \otimes (1,0)_B \succ_{\v{\gamma}_S\otimes\v{\gamma}_B} \v{\gamma}_S\otimes (\epsilon,1-\epsilon)_B,
\end{equation}
where $\v{p}_S$ denotes the initial energy populations of $S$ and the battery $B$ is a two-level system with energy splitting~$W$ and $\succ_{\v{\gamma}_S\otimes\v{\gamma}_B}$ denotes thermomajorization~\cite{horodecki2013fundamental}. Such minimal $\epsilon$ can be obtained by solving a corresponding set of inequalities. 

However, when one is limited to Markovian interactions with the bath, the optimal error $\epsilon_{\mathrm{MTP}}(W)$ is determined by \emph{continuous thermomajorization} (see the accompanying paper \cite{lostaglio2022continuous}). In practice, the optimal error can be computed using our algorithm. To do so, one can take a sequence of increasing $W$ and, for each of them, look for the smallest $\epsilon$ such that the transformation \begin{equation}
	\v{p}_S \otimes (1,0)_B \stackrel{\textrm{MTP}}{\longmapsto} \v{\gamma}_S\otimes (\epsilon,1-\epsilon)_B,
\end{equation}
is possible. The algorithm returns a yes/no answer concerning the existence of an MTP for fixed $(W,\epsilon)$. The minimal $\epsilon$ such that an MTP exists can be achieved with exponentially good approximation by a sequence of systematic guesses for $\epsilon$. In this way, one constructs the plots presented in the main text.

\emph{Optimal cooling of a four-level system.} In this application we considered a protocol that starts from a thermal state with population vector $\v{\gamma}=(\gamma_0,\gamma_1,\gamma_2,\gamma_3)$, where $\gamma_k = e^{-\beta k\Delta }/(\sum_k e^{-\beta k \Delta })$. Then, a unitary $U$ induces a full population inversion, meaning that the populations are rearranged in a non-decreasing order:
\begin{equation}
	\v{\gamma} \stackrel{U}{\longmapsto} \v{\gamma}^\uparrow = (\gamma_3,\gamma_2,\gamma_1,\gamma_0). 
\end{equation}
Intuitively, this step pumps free energy into the system, making it as much out-of-equilibrium as possible. From here, we would like to apply the MTP that achieves the highest possible ground state population. To do so, we use our algorithm to compute, for every fixed $\beta$, the full set of states achievable from $\v{\gamma}^\uparrow$ by an arbitrary MTP. From these states, we select the one with the highest ground state population. In this way, we constructed the plot presented in the main text.

\begin{figure}
	\includegraphics[width=0.45\columnwidth]{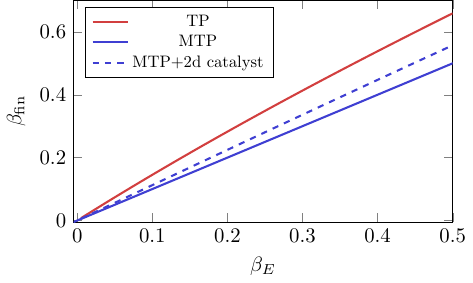}
	\caption{\label{fig:catalyst}\textbf{Optimal cooling of a two-level system with catalyst.} Optimal final inverse temperature $\beta_{\mathrm{fin}}$ as a function of the environmental inverse temperature $\beta_E$ for a two-level system initially at inverse temperature $\beta_E/2$. We present the optimal cooling achievable under thermal processes (red, TP), Markovian thermal processes (blue, MTP) and Markovian thermal process with an additional thermal catalyst of dimension 2 (dashed blue).}
\end{figure}

\emph{Catalysis.} In the case of catalysis, we considered the transformation 
\begin{equation}
	\v{\gamma}_{\beta_E/2} \otimes \v{c} \stackrel{ \textrm{MTP}}{\longmapsto} 	\v{\gamma}_{\beta_{\mathrm{fin}}} \otimes \v{c},
\end{equation}
where 
\begin{equation}
	\v{\gamma}_x:= \left(\frac{1}{1+e^{-x}}, \frac{e^{-x}}{1+e^{-x}} \right),	
\end{equation}
setting energy units to~$1$ and $\v{c} = \v{\gamma}_\beta$ (a single qubit thermal catalyst). The system starts at twice the temperature of the environment (inverse temperature $\beta_E/2$). For each fixed $\beta_E$, we looked for the highest $\beta_{\mathrm{fin}}$ for which the above transformation is possible, by applying our algortihm to a sequence of systematic guesses for $\beta_{\mathrm{fin}}$. One can also compute the highest achievable $\beta_{\mathrm{fin}}$ for any given $\beta_E$ using thermomajorization. The results are summarized in Fig~\ref{fig:catalyst}. As one can see, while under MTP one cannot cool to temperatures lower than that of the environment, this becomes possible with a single qubit thermal catalyst. For the sake of comparison, we also plotted the minimal temperature achievable under thermal processes (TP), which can access arbitrary non-Markovian effects (TP satisfy only properties~P2~and~P3 in the main text).

\emph{Efficiency of bio-molecular processes.} Photoisomerisation is a process by which a molecule is excited to a new configuration, triggering fundamental processes in biological systems. A simple model of a photoisomer was used in Ref.~\cite{halpern2020fundamental}, whereby the system Hamiltonian is parametrized by the relative rotation $\varphi$ between two molecular components around a double bond. For each $\varphi$, one considers two possible energy states $E_0(\varphi)$ (`ground') and $E_1(\varphi)$ (`excited'). The $\varphi = 0$ configuration is known as \emph{trans} and the $\varphi= \pi$ as \emph{cis}. In Ref.~\cite{spaventa2021non}, the system is then reduced to the three-level system $\{E_0(0), E_0(\pi), E_1(0)\}$, where $E_0(0) \leq E_0(\pi) \leq E_1(0)$ (models involving further levels can however be analysed with our algorithm). The photoisomer, initially in a thermal state of the trans configuration, is excited by a light source at $t=0$ to a trans state $\v{p}(0) = (p_{00},0,p_{10})$, where $p_{10} = 1-p_{00}$ denotes the degree of excitation. By means of rethermalization with an environment, the photoisomer reaches a new state $\v{q} = (q_{00},q_{0\pi},q_{10})$ at some intermediate time, before rethermalizing completely. The \emph{photoisomerization yield} is defined as the maximum achievable $q_{0 \pi}$, i.e. the highest possible probability of switching to a cis configuration as a function of the other parameters (initial excitation probability, gaps, temperature). In Ref.~\cite{halpern2020fundamental}, $q_{0 \pi}$ was optimized over all thermal processes, obtaining a yield $\gamma_{\mathrm{TP}}$. In Ref.~\cite{spaventa2021non}, the yield $\gamma_{\mathrm{MTP}}$ was computed by maximising over all Markovian thermal processes, using a preliminary version of our results~\cite{preliminary}. The presence of a gap $\gamma_{\mathrm{TP}} > \gamma_{\mathrm{MTP}}$ has been put forward as model-independent, quantitative evidence that environmental memory effects can significantly increase the efficiency of bio-molecular processes~\cite{spaventa2021non}.

\end{document}